\documentclass[%
nofootinbib,
superscriptaddress,
 preprint,
 amsmath,amssymb,
 aps,
 prl,
 longbibliography,
 floatfix,
 lengthcheck,%
]{revtex4-1}

\usepackage{hyperref}
\usepackage[hyphenbreaks]{breakurl}
\usepackage{RabitzGate_SCI}

\hypersetup{
  colorlinks=true,
  linkcolor=blue!50!red,
  urlcolor=red!70!black
}

\usepackage{soul}
\newcommand{\citeurl}[1]{See {\burl{#1}}.}

\begin{document}

\title{Theory of quantum control landscapes: Overlooked hidden cracks}

\author{Dmitry V. Zhdanov}
\email{dm.zhdanov@gmail.com}
\affiliation{Northwestern University, Evanston, Illinois 60208, USA}

\begin{abstract}
\epigraph{Skepticism is a normal and healthy attitude in science, as 
opposed to religion, and it is for the believer to give a convincing proof that the 
anticipated miracle is about to happen.}{\href{https://www.coulomb.univ-montp2.fr/user/michel.dyakonov?lang=en}{M. I. Dyakonov}, \cite{2014-Dyakonov}}
Why does controlling quantum phenomena appear easy to achieve? Why do effective quantum controls appear easy to find? Why is chemical synthesis and property optimization easier than expected? How to explain the commonalities across the optimal control applications in quantum mechanics, chemistry, material science, biological evolution and engineering? The theory of quantum control landscapes (QCL) is developed by Prof. Rabitz and his colleagues to address these puzzling questions. Unfortunately, the obtained conclusions are subject to  misinterpretations which are spread in hundreds of published papers. We investigate, summarize and report several previously unknown subtleties of the QCL theory which have far-reaching implications for nearly all practical applications.
\end{abstract}

\maketitle 

Late 1990s and early 2000s were the years of triumphant success of the quantum feedback loop experiments enabling optimal control of ultra-small systems, such as atoms and molecules \cite{1998-Assion,2000-Bartels,2001-Brixner}. These experiments started the new era of using lasers in chemical analysis, nanotechnology and quantum information science. The theory required to support these novel applications had already been well-developed in the early 1980s (see e.g. \cite{1990-Butkovskiy}). Nevertheless, it was still tempting to find a formalism best tooled for specifics of optimal control in quantum-mechanical realm. QCL arose as a response to this challenge. 

Nowadays, QCL is a mature theory covered in a number of reviews \cite{2010-Brif,2011-Brif}. Nevertheless, we will show that this theory has a number of broadly overlooked pitfalls leading to profound practical implications. Some of these pitfalls were reported before but noticed only by a very small circle of specialists. Others will be reported here for the first time. By critically reviewing the history of QCL research, we will identify and summarize those accumulated mistakes and incorrect interpretations that continue to proliferate in today's scientific literature. We will begin with outlining a typical optimal quantum control experiment to clarify the actual practical questions and challenges. Then we will discuss how the successes and failures in addressing these challenges had shaped the core paradigms of QCL theory over the years. Based on results of this analysis, a way to assess the prospective future of QCL will be proposed.

We will restrict our discussion to canonical quantum control task of bringing
 the system into desired quantum state $\ket{f}$ at given time $T$. This way, one could initiate photochemical reactions (e.g., change the state of molecular switch), conduct ultrafast spectroscopy studies or initialize a quantum register \cite{2012-Hoff}. The most established technology to do such things in laboratory is \emph{coherent control} (CC)\cite{2007-Nuernberger,2010-Brif}. CC originated in the early 1960s with the first NMR experiments and flourished in the 1990s after the emergence of compact and affordable femtosecond lasers. CC is based on exposing the system to a series of microwave or laser electromagnetic pulses $\EE(t)$. In a typical experiment, the laser output $\EE_0(t)$ is first transmitted through a \emph{pulse shaper}, the device which splits $\EE_0(t)$ into $K/2$ spectral components, and then allows us to adjust the amplitude and phase for each of them -- so, $K$ controlled parameters $\eee{=}{\ee_1,...,\ee_{2K}}$ in total. With modern broadband lasers and pulse shapers with $K{\simeq}10^2{-}10^3$ one is capable to prepare any desirable profile $\EE(t)$. However, what is the practical power of this capability? Specifically:
\begin{enumerate}[(a)]
\item \label{_qn_1}Is it possible to reach the desired $\ket{f}$? (Is system controllable?)
\item \label{_qn_2}If so, how difficult is it to find the appropriate parameters $\eee$?
\end{enumerate}
Let us clarify what is exactly meant by controllability. The evolution of closed quantum system from its initial state $\ket{\Psi_0}$ at $t{=}0$ to a final state $\ket{\Psi_T}$ at $t{=}T$ is always governed by a certain unitary operator $\hat U(\eee)$:
\begin{gather}\label{__evo}
\ket{\Psi_T}{=}\hat U(\eee)\ket{\Psi_0}.
\end{gather}   
\begin{definition}
The system is said to be controllable if for any randomly chosen $N{\times}N$ unitary matrix $\hat U'$ where exist at least one policy $\{\eee,T\}$ such that $\hat U(\eee){=}\hat U'$. Thence, controllability implies that an arbitrary final state $\ket{\Psi_T}$ can be obtained from an arbitrary initial one $\ket{\Psi_0}$.
\end{definition}

A quite generic answer to the question \eqref{_qn_1} was found in early 70s by Jurdjevic and Sussmann \cite{1972-Jurdjevic}:
\begin{theorem}
For any closed $N$-level quantum system ($N{<}\infty$) satisfying certain well-defined and physically mild assumptions there exists a time $T_0{<}\infty$ such that the system is controllable if $T{>}T_0$.
\end{theorem}

The notion of quantum control landscapes (QCL) was introduced by H. Rabitz to formalize the  question \eqref{_qn_2} in the ideal limit of very large $K$. Let us introduce so-called \emph{performance index}
\begin{gather}\label{__pi}
\JJ{=}|\scpr{\Psi_{T}}{f}|^2.
\end{gather}
The index $\JJ$ characterizes quality of the chosen control policy $\{\eee, T\}$. The maximal value $J{=}1$ indicates that we have achieved exactly what we wanted. The smaller $J$, the larger deviation of $\ket{\Psi_{T}(\eee)}$ from $\ket{f}$. The multiparameter function $J(\eee)$\footnote{Strictly speaking, $J(\eee)$ is a functional in the limit $K\to\infty$.} is called \emph{quantum control landscape} (QCL).

\begin{figure}[tbp]
\centering\includegraphics[width=\columnwidth] {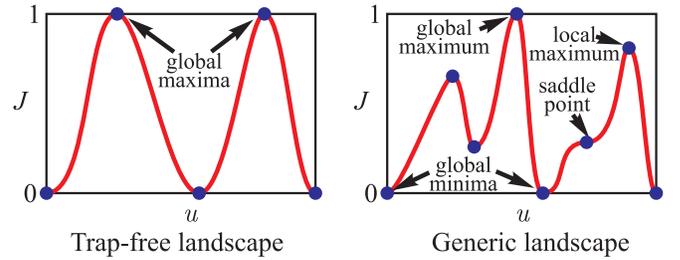}
\caption{Sample 1D trap-free and generic landscapes featuring various types of critical points. Critical point $\eeecr$ is called maximum (minimum) if $J(\eee)\leq J(\eeecr)$ ($J(\eee)\geq J(\eeecr$) for all $\eee'$ in some $\epsilon$-neighborhood of $\eeecr$. The point $\eeecr$ is called saddle point if it is neither minimum nor maximum, but $\der{J}{\ee_i}{=}0$ for any $i{\in}\{1,...,K\}$.
\label{@FIG.02}}
\end{figure}

The answer to question \eqref{_qn_2} depends on the properties of QCL. As any function, $J(\eee)$ in principle may have a variety of critical points: global and local minima and maxima as well as saddle points (see Fig.~\ref{@FIG.02}). An ``easy'' landscape does not contain any saddle points or local maxima (also referred as ``traps''). In this case, the optimal policy can be determined by simply ``climbing'' to the top of any of the landscape's peaks. This can be done by gradually adjusting arbitrary initial controls $\eee$ using any local gradient accent algorithm. However, this recipe will not work in the generic case when local maxima and saddles are also present. In this ``difficult'' case, much more involving global search algorithms are needed to identify the highest peak(s).

In 2004 Rabitz published the work \cite{2004-Rabitz} which became foundational for QCL theory. As of today, this work is \emph{de facto} the iconic and second most cited reference of QCL theory. The central  result is the claimed rigorous proof of the following statement:
\begin{wrong_theorem}\label{__theo_no_traps}
QCL of controllable systems are trap-free: all critical points are global extrema.
\end{wrong_theorem}
To prove the theorem, the authors consider $\JJ(\eee)$ as a compound function $\JJ(\hat U(\eee))$ (where $\hat U(\eee)$ is defined by \eqref{__evo}) and make the following two claims.
\begin{wrong_proposition}\label{__prop_fr}
If system is controllable then for any $\eee$ all unitary operators $\hat U'$ which are close to $\hat U(\eee)$ (i.e., $\hat U'{=}\hat U(\eee){+}\epsilon\hat{\delta U}$) can be obtained via small variation of controls $\hat U'{=}\hat U(\eee{+}\epsilon\ddee{+}o(\epsilon))$.
\end{wrong_proposition}
\begin{wrong_proposition}\label{__prop_kin}
The index $\JJ(\hat U)$ considered as a function of the set of all unitary operators $\hat U$ has only global minima and maxima.
\end{wrong_proposition}

\begin{figure}[tbp]
\centering\includegraphics[width=0.8\columnwidth] {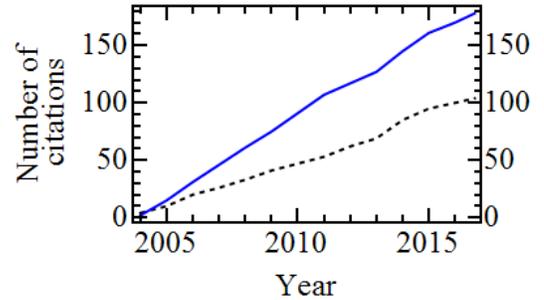}
\caption{Impact of the paper \cite{2004-Rabitz}. Cumulated number of self-citations (black, dashed) and independent citations (blue, solid). Data source: Google Scholar (collected September 23, 2017).\label{@FIG.01}}
\end{figure}

Both propositions \ref{__prop_fr} and \ref{__prop_kin} are wrong. For proposition \ref{__prop_kin} this can be seen from direct calculation of the first variation of $J$ using eqs.~\eqref{__evo} and \eqref{__pi}:
\begin{gather}\notag
\delta\JJ{=}\matel{\Psi_{0}}{\hat U}{f}\matel{\Psi_{0}}{\hat{\delta U}}{f}^*{+}c.c.
\end{gather}
It is obvious that $\delta\JJ{=}0$ for any $\hat{\delta U}$ such that $\matel{\Psi_{0}}{\hat U}{f}{=}0$. More detailed analysis shows that all such points are saddles of $\JJ(\hat U)$. Proposition \ref{__prop_fr} was first demonstrated to be incorrect in numerical experiments as early as in 2007 \cite{2007-Chakrabarti} and then rigorously disproved in the works by De Fouquieres and Schirmer \cite{2013-De_Fouquieres} as well as Pechen and Tannor \cite{2011-Pechen}.

Unfortunately, up to now, the work \cite{2004-Rabitz} is still generally considered valid proof of ``easiness'' of quantum landscapes (see Fig.~\ref{@FIG.01}), even by the researchers who are aware about the works \cite{2011-Pechen,2013-De_Fouquieres} (e.g., \cite{2013-Palao,2015-Anderson,2016-Liebermann}). The controversies are further amplified by the fact that Rabitz treats such misleading citations merely as an evidence that ``the importance of the landscape topology to determining the feasibility of quantum control is beginning to be more widely recognized'' (citation of the work \cite{2009-Merkel} in the Rabitz's paper \cite{2011-Moore}). Furthermore, he did not retract his original claim regarding ``easiness'' of quantum landscapes. Instead, he relaxed it to the following form \cite{2012-Moore}:
\begin{quote}
The broad success of optimally controlling quantum systems with external fields has been attributed to the favorable topology of the underlying control landscape, where the landscape is the physical observable as a function of the controls. The control landscape can be shown to contain no suboptimal trapping extrema upon satisfaction of reasonable physical assumptions, but this topological analysis does not hold when significant constraints are placed on the control resources.
\end{quote}

More explicitly, the following changes were made: 
\begin{enumerate}[\{i\}]
\item \label{_theo_r_i} A ``trap-free control landscape'' is now allowed to include a saddle points. 
\item \label{_theo_r_ii} Correspondingly, the fact that proposition \ref{__prop_kin} is wrong is admitted but not considered as a big deal.
\item \label{_theo_r_iii} Proposition \ref{__prop_fr} (also in the relaxed form admitting for saddles) is now simply \emph{postulated} by appealing to laboratory results and numerical experiments. Specifically, the function $\hat U(\eee)$ is \emph{believed} to almost always be trap-free (that is, \emph{locally controllable}) provided that we have enough control resources (i.e., that $K$ is large enough). The latter is treated as a ``reasonable physical assumption''.
\end{enumerate}

\begin{figure}[tbp]
\centering\includegraphics[width=0.9\columnwidth] {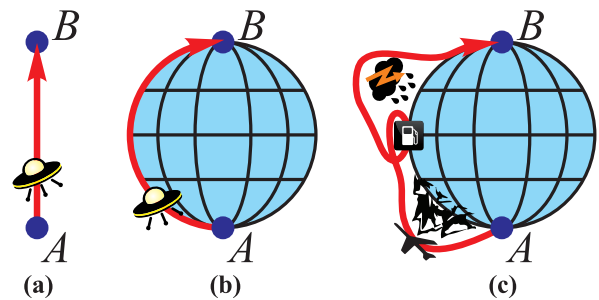}
\caption{Classical air travel analog of quantum optimal control problem.
\label{@FIG.03}}
\end{figure}

To understand the physical meaning of this surrogate of the theorem \ref{__theo_no_traps}, let us draw a simple analogy. Imagine that we have a ``flying saucer,'' a fully controllable apparatus which can take off and land at any point and move equally well everywhere in the space regardless of presence/absence of air, weather conditions etc. We are looking for the easiest way to reach a point $B$ from point $A$. If we would address this question to a little boy or an ancient Egyptian, we would be advised to go straight from $A$ to $B$, as shown in Fig.~\ref{@FIG.03}(a). However, we know now that the Earth is round! And, yes, we have a flying saucer, but not an earthmoving machine. So, if $A$ and $B$ are located at the South and North poles then our best option is the trajectory shown in Fig.~\ref{@FIG.03}(b). Note that upon launch we have to fly in the direction orthogonal to our destination! 

Violation of the proposition \ref{__prop_kin} implies that we always have the case of Fig.~\ref{@FIG.03}(b) when ``climbing'' QCL! 
And the quantum ``earthmoving machines'' are strictly forbidden by laws of quantum mechanics! No matter how much control resources we have.

Furthermore, in reality we deal with an optical modulators, i.e., a physical aircrafts rather than fictitious flying saucers. And in order to successfully accomplish the trip from $A$ to $B$ our engineers, pilots and dispatchers need the specific instructions. In the case of crush we should be able to investigate the reason: e.g., weather condition, lack of fuel, collision with a cow when flying at low altitude, etc. The original (incorrect) theorem \ref{__theo_no_traps} proposed controllability as a well-defined and rigorous test for assessing ``quantum aircrafts'' and investigating such an accidents. On contrary, the updated surrogate version requires us to have ``sufficient control resources''. The practical value of this requirement is reminiscent to the following old Russian joke:
\begin{quote}
The story goes that during World War II
an inventor appeared with an idea of extreme military
value 
 and insisted that they take him to the
very top, that is, to Stalin.
 
``So, tell me, what is it about?''

``It's simple, comrade Stalin. You will have three
buttons on your desk, a green one, a blue one, and a
white one. If you press the green button, all the enemy
ground forces will be destroyed. If you push the blue
button, the enemy navy will be destroyed. If you push
the white button, the enemy air force will be destroyed.'' 

``OK, sounds nice, but how will it work?''

``Well, it's up to your engineers to figure it out! I
just give you the idea.''
\end{quote}

The absence of a strict definition of ``sufficient control resources'' in principle allows one to unfalsifiably treat any experimental evidence of trap-free landscapes as a confirmation of the argument \{\ref{_theo_r_iii}\}, such as was done, e.g., in Ref.~\cite{2011-Moore}. However, such interpretation may not be fully correct. For example, the ``simplicity'' of the experimental multiparameter QCL may result from statistical effects, such as asymptotic aggregation (see, e.g., \cite{2007-Boss}, Sec 11). We also should keep in mind that the requirement of ``sufficient control resources'' is not always a ``reasonable physical assumption''. The realistic ``easiest'' trajectory in our aircraft example might look like shown in Fig.~\ref{@FIG.03}(c). It will be essentially defined by the technical capabilities of the chosen aircraft, weather conditions, international regulations etc. In that sense, quantum control problems are no different. The parameters $\ee_i$ of our pulse shaper are always bounded: $\ee_i{\in}[\ee_{i,\idx{min}},\ee_{i,\idx{max}}]$. There are uncontrollable sources of noise and decoherence which define the ``weather conditions''. The maximal intensities and laser spectral ranges can also be constrained by ``non-damaging regulations''. It seems evident and also can be rigorously proven that the systems under such constraints are normally not fully controllable (see, e.g., \cite{2003-Altafini}), so that any questions about their QCL are meaningless without a detailed account for all above technical information \cite{2015-Zhdanov,2017-Zhdanov}. 

Despite of all these issues, the arguments of a type \{\ref{_theo_r_i}\}-\{\ref{_theo_r_iii}\} were claimed to explain a broad variety of experimental evidences of trap-free landscapes, even in the areas far beyond the scope of quantum control, such as chemical synthesis, property optimization etc. (see, e.g., Ref.~\cite{2011-Moore}). In his recent grant proposal\footnote{\citeurl {https://www.templeton.org/grant/universal-operating-principle-for-optimal-control-in-the-sciences-optisci-over-vast-length-and-time-scales}
}
 and paper \cite{2017-Russell2}, Rabitz even intends to apply such arguments to explain the biological evolution! The theoretical justification for such a revolutionary extension of the theory is given in the paper \cite{2017-Russell}. Its primary result is the generic proof that the sufficient control resources requested in the argument \{\ref{_theo_r_iii}\} are very modest for nearly all kinds of controllable systems. The formal claim (theorem 4.2 in \cite{2017-Russell}) sounds a bit horrifying and can be split into the following two theorems.

\begin{theorem}\label{__theo_rsl_a}\footnote{The justification of theorem \ref{__theo_rsl_a} provided in \cite{2017-Russell} is also incorrect. However, we will avoid discussing this issue here for brevity.}
For any $N$-dimensional closed quantum system and substantially large natural number $Z$, it is possible to formally introduce a set $\eee$ of bounded controls $\ee_{j}{\in}[-\kappa,\kappa]$, $j{=}1,...,(N^2{-}1)Z$ such that 1) the system is controllable, and 2)
its landscape $J(\eee)$ is trap-free.
\end{theorem}

\begin{wrong_theorem}\label{__theo_rsl_b}
For a controllable system with trap-free landscape $J(\eee)$ fixing any single control parameter $\ee_{j'}{=}c$ may introduce local maxima and minima into the new control landscape $J(\eee|\ee_{j'}{=}c)$ (a function of the remaining variables $\ee_{j{\ne}j'}$ only) only for a null set of values of $c{\in}\mathbb{R}$.
\end{wrong_theorem}

The net idea of applying these theorems is following. We begin with introducing a very rich set of controls $\eee$ satisfying theorem \ref{__theo_rsl_a} and, thus, also the assumption \{\ref{_theo_r_iii}\}. Then, we start ``freezing'' controls $\ee_j$ one-by-one. Theorem \ref{__theo_rsl_b} implies that the probability of creating a trap by any such control elimination is nearly zero. Hence, the iterative eliminations can be repeated until they start compromising the system controllability. Thus, we can ``freeze'' most of controls $\ee_{j}$ at arbitrary values without destroying the trap-free QCL structure. 

Let us show that theorem \ref{__theo_rsl_b} is wrong. Its proof proposed in \cite{2017-Russell} relies on so-called parametric transversality (PT) theorem (see, e.g., \cite{1985-Arnol'd}, p.~39, Lemma 1). In context of our problem, the core idea of PT theorem can be expressed as follows.
\begin{theorem}
\label{__theo_pt}
Let $J(\eee)$ be a trap-free landscape having no local minima. Then, for any given control index $j$ and real number $J_0$ there may be only a null set $\manC$ of values $c$ for which the constrained landscape $J(\eee|\ee_j{=}c)$ includes such points $\eee'$ that: 1) $\eee'$ is a local maximum or minimum of $J(\eee|\ee_j{=}c)$, and 2) $J(\eee'|\ee_j{=}c){=}J_0$.
\end{theorem}

\begin{figure}[tbp]
\centering\includegraphics[width=0.65\columnwidth] {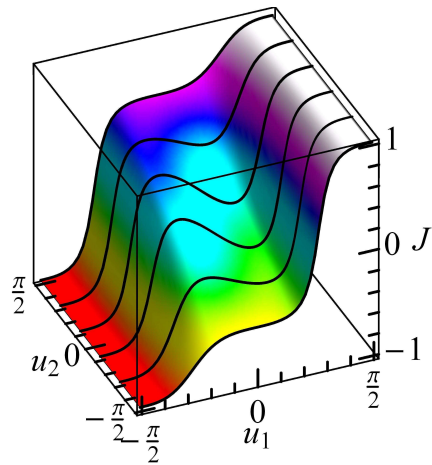}
\caption{The example landscape $J(\ee_1,\ee_2)$ illustrating the incorrectness of theorem \ref{__theo_rsl_b}. The solid black lines indicate the constrained landscapes $J(\ee_1,\ee_2{=}c)$ for different values of $c$.
\label{@FIG.04}}
\end{figure}

The authors derive theorem \ref{__theo_rsl_b} by claiming that theorem \ref{__theo_pt} implies that the constrained landscape $J(\eee|\ee_j{=}c)$ is trap-free for almost all values of $c$. However, the claim is incorrect. This can be demonstrated by a simple counterexample involving just two control parameters $\ee_{1,2}{\in}[{-}\frac{\pi}2,\frac{\pi}2]$. Consider function $J(\ee_1,\ee_2)=\frac2{\pi}(\tan(\ee_1)^3{-}\tan(\ee_1)\cos(\ee_2){+}\tan(\frac{\ee_2}{2}))$ shown in Fig.~\ref{@FIG.04}. This function has no local minima or maxima. Moreover, even its range does not change after fixing the argument $\ee_2$: $J(\ee_1,c){\in}[-1,1]$ for any $c{\in}[{-}\frac{\pi}2,\frac{\pi}2]$. Theorem \ref{__theo_pt} is satisfied for $j{=}2$ and any $J_0{\in}(-1,1)$. Indeed, for any $J_0$ the set $\manC$ consists of at most two isolated points $c{\in}({-}\frac{\pi}2,\frac{\pi}2)$. If the authors' claim is also correct then ``freezing'' the parameter $\ee_2$ at arbitrary value $c{\in}[{-}\frac{\pi}2,\frac{\pi}2]$ would imply nearly zero risk of having a trap in the resulting constrained landscape $J(\ee_1,c)$. However, Fig.~\ref{@FIG.04} shows that such landscapes $J(\ee_1,c)$ contain both local minimum and maximum for \emph{any} $c{\in}(-\frac{\pi}2,\frac{\pi}2)$.

We conclude noticing that the history of QCL in many senses resembles the attempts to build a quantum Turing machine. In both cases there is a huge gap between theory and experiment, ambitious anticipations but modest outcomes. Actually, the joke about universal military transformation device was quoted from the excellent paper \cite{2003-Dyakonov} by Dyakonov published in early 2003. Dyakonov used it to express a very similar skepticism about building universal quantum computer. All his objections were proved by time and remain actual today. In particular, he properly predicted that even the seemingly very simple task of realizing full Shor's algorithm for 4-bit numbers (or even simply producing a technical manual) will remain unsolvable. Inspired by Dyakonov's experiment, we will also formulate a realistic practical challenge for QCL theory in order to assess the progress in, say, 10 years. We definitely do not want to challenge the theory by complicated questions about biological evolution. Let us return back to our simplest original problem with a pulse shaper. Suppose I have a molecule with reduced $N$-level model Hamiltonian $\hat H$ subject to radiational decay described by Markovian Liouvillian ${\cal L}_{\idx{rad}}$ (sorry, the laws of physics do not allow to switch it off). The molecule can interact with laser pulse via electrodipole interaction term ${-}\vec{\hat d}\vec\EE$. I also provide a complete technical specifications of my laser and pulse shaper. I want to know whether I have enough control resources to enjoy trap-free QCL for a given control time $T$. No demagoguery and philosophy, please.

\bibliographystyle{apsrev4-1}
\nocite{apsrev41Control}
\bibliography{RabitzGate_SCI}

\end{document}